\documentstyle[tighten,preprint,aps,prd]{revtex}

\begin{document}
\draft

\def\Bbb{\bf}
\def\BbbR{{\Bbb R}}
\def\BbbZ{{\Bbb Z}}
\def\a{\alpha}
\def\b{\beta}
\def\d{\delta}
\def\D{\Delta}
\def\e{\epsilon}
\def\g{\gamma}
\def\m{\mu}
\def\n{\nu}
\def\ni{\noindent}
\def\no{\nonumber}
\def\o{\omega}
\def\O{\Omega}
\def\S{\Sigma}
\def\t{\theta}
\def\T{\Theta}
\def\be{\begin{equation}}
\def\ee{\end{equation}}
\def\RPthree{{\BbbR P^3}}
\def\arsinh{\mathop{\rm arsinh}\nolimits}

\preprint{\vbox{\baselineskip=12pt
\rightline{IUCAA--34/98}
\rightline{}
\rightline{gr-qc/9906063}
}}
\title{On the Brown-York quasilocal energy, gravitational
charge, and black hole horizons}
\author{
Sukanta Bose\footnote{Electronic address:
{\em sbose@iucaa.ernet.in}}
and
Naresh Dadhich\footnote{Electronic address:
{\em nkd@iucaa.ernet.in}}
}
\address{IUCAA, Post Bag 4, Ganeshkhind, Pune 411007, India}
\date{September 1998}

\maketitle
\parshape=1 0.75in 5.5in

\maketitle
\begin{abstract}%
 
We study a recently proposed horizon defining identity for certain black 
hole spacetimes. It relates the difference of the Brown-York quasilocal energy
and the Komar charge at the horizon to the total energy of the spacetime. The 
Brown-York quasilocal energy is evaluated for some specific choices of 
spacetime foliations. With a certain condition imposed on the matter 
distribution, we prove this identity for spherically symmetric static black 
hole solutions of general relativity. For these cases, we show that the 
identity can be derived from a Gauss-Codacci condition that any 
three-dimensional timelike boundary embedded around the hole must obey. We also
demonstrate the validity of the identity in other cases by explicitly applying 
it to several static, non-static, asymptotically flat, and asymptotically 
non-flat black hole solutions. These include the asymptotically 
Friedmann-Robertson-Walker (FRW) solutions and the case of a black hole with a 
global monopole charge. 

\end{abstract}  

\narrowtext
\bigskip

The concept of a black hole horizon, ever since its birth with Lemaitre's 
demonstration of the non-singularity of the Schwarzschild horizon, has 
played a monumental role in the understanding of the causal structure 
of several spacetimes. It has been associated with many general relativistic 
theorems and laws of import, e.g., the singularity theorems \cite{Pen1}, the 
black hole area theorem \cite{Hawk1}, and the laws of black hole 
mechanics. It plays an important role in Hawking's semiclassical calculation 
on the evaporation of a black hole \cite{Hawkevap} and is also related to its 
entropy \cite{Bekent}. Finally, although the horizon may not have any 
special significance in the frame of a freely falling observer, to an 
asymptotic inertial observer it behaves very much like a physical membrane 
(see, e.g., Ref. \cite{Thorne}). 

The behavior of families of null geodesics in a given spacetime determines if
a horizon exists in it. Moreover, it is known that the properties of such a 
family are affected by the matter stress-tensor through the Raychaudhuri 
equation \cite{Raychau}. This prompts one to ask if there exists a direct 
characterization of the black hole horizon in terms of the quasilocal energy 
or gravitational ``charge'' of bounded regions embedded in such spacetimes. To 
seek such a characterization will be the main aim of this paper. 

It is well known that there are inherent difficulties in defining energy in 
general relativity (GR), essentially owing to its non-localizability. So far, 
considerable effort has been put in to formulate a satisfactory definition. 
For quasilocal energy, we  shall here adopt the definition given by Brown 
and York \cite{BY} (henceforth referred to as BY), which can be summarized
as follows: The system under consideration is a spatial three-surface $\Sigma$
bounded by a two-surface ${\sf B}$ in a spacetime region that can be 
decomposed as a product of a spatial three-surface and a real line-interval 
representing time. The time evolution of the two-surface boundary ${\sf B}$ is 
the timelike three-surface boundary ${}^3 {\sf B}$. When $\S$ is taken to 
intersect ${}^3 {\sf B}$ orthogonally, the BY quasilocal energy is defined as:
\be \label{BYE}
 E = \frac{1}{8 \pi} \oint_{\sf B} d^2 x \sqrt{\sigma} (K - K_0) \ \ , 
\ee
where $\sigma$ is the determinant of the 2-metric on ${\sf B}$, $K$ is the  
trace of the extrinsic curvature of ${\sf B}$, and $K_0$ is a reference term 
that is used to normalize the energy with respect to a reference spacetime, not
necessarily flat. (Here, we employ geometrized units, with $G=1=c$.) Henceforth
we will use the acronym ``QLE'' to denote the BY quasilocal energy.  

To compute the QLE for asymptotically flat solutions, we will choose the 
reference spacetime to be Minkowski. In that case, $K_0$ is the trace of the 
extrinsic curvature of a two-dimensional surface embedded in flat spacetime, 
such that it is isometric to ${\sf B}$. For asymptotically 
non-flat solutions, the Hawking-Horowitz prescription \cite{HH,BL} will be 
employed to determine the reference spacetime.

For describing interactions other than gravity, one uses separate measures 
for the mass (or energy) of a  particle and its charge (which determines its 
strength of coupling to the field). By  contrast, in GR there is a single 
measure for both. Besides, the gravitational field of a particle itself 
contributes to its gravitational charge since it contains energy. It is this 
aspect of the gravitational field that has been the main cause of ambiguity in 
attempts at distinguishing charge from energy in GR. However, one possibility 
for defining gravitational charge would be to suitably adopt the Gauss law in 
Newtonian gravity to the case of GR \cite{DC}. For asymptotically flat, static 
spacetimes, this generalization is readily implemented: Let $t^a$ be a 
timelike Killing vector field normalized in such a way that its norm approaches
unity at infinity. Then, for an isolated system, the total mass $M$ within a
two-surface ${\sf B}$ is the total force that must be exerted by an observer at
infinity on a test matter with unit surface mass density spread on ${\sf B}$, 
such that each point on ${\sf B}$ follows an orbit of a Killing vector such as
$t^a$. Then,
\be \label{Mc}
M_c = \frac{1}{4 \pi} \oint_{\sf B} {\bf g}. {\bf ds} \ \ ,
\ee
where ${\bf g}=-N\nabla (\ln N)$, $N= \sqrt{-t^a t_a}$, and the integral is
taken  over the closed 2-surface ${\sf B}$. The norm of ${\bf g}$ evaluated at 
the horizon  gives  the surface gravity of the hole. This  is  how 
gravitational charge is intimately related to surface gravity (and, therefore, 
the temperature)  of a black hole. Note that, for a suitable choice of
foliation of the spacetime region of interest with spacelike hypersurfaces, one
can choose a time coordinate such that $N$ plays the role of lapse function. 
This is the choice that we will exercise below. 

The above expression can be equivalently written as 
\be \label{Komar}
M_c = -\frac{1}{8 \pi} \oint_{\sf B} \epsilon_{abcd} \nabla^c  t^b 
\ \ ,\ee
where $\e_{abcd}$ is the volume element on the spacetime. It is in fact the 
Komar mass \cite{Komar}.
Equation (\ref{Komar}) defines a conserved (both in space and 
time) gravitational charge when the spacetime is vacuum and admits a timelike 
Killing vector. When either of these conditions is relaxed, $M_c$ depends upon
the location of ${\sf B}$. In such a case it describes the {\em quasilocal}
charge associated with the spatial volume bounded by ${\sf B}$. 

It is clear from the above definition of gravitational charge that it is the 
lapse function that determines it. On the other hand QLE is not 
at all sensitive to it and is instead determined entirely by the spatial 
metric. Hence, the measures of quasilocal charge and energy will, in general, 
be different. In what follows, by the energy and charge of a spacetime region, 
we shall mean the gravitational quasilocal energy and charge (references to the
electric or magnetic charge of that region will be made explicitly).
 
Dadhich \cite{Da1,Da2} has recently proposed a novel energetics 
characterization of the horizon of a black hole in asymptotically flat, 
spherically symmetric static (SSS) spacetimes. He proposed that its location 
is at that curvature-radius, $r$, at which the following identity holds:
\be \label{Ident}
E_{\cal H} -E_\infty = M_{\cal H} \ \ ,
\ee
where $E_{\cal H}$ is the QLE at the horizon, $E_\infty$ is the total energy of
the spacetime, and $M_{\cal H}$ is the value of the gravitational charge, 
$M_c$, at the horizon. The physical interpretation of this identity is as 
follows. The gravitational charge essentially defines the strength of the 
leading Newtonian potential. In the Newtonian sense, therefore, it measures the
strength of the gravitational pull exerted by a body. On the other hand, as 
shown by Eq. (\ref{BYE}), the field energy, $E_\infty -E_{\cal H}$, is obtained
completely from the spatial part of the metric. Its significance can be seen by
noting that it is related to the spatial conformal factor (see, e.g., Ref. 
\cite{Will}) that arises in higher post-Newtonian orders in the expansion of 
the spacetime metric around a non-rotating matter distribution. It, therefore,
measures the amount of ``curvature of space'' due to that distribution in the 
sense that the spatial components of the Riemann curvature tensor (up to 
post-Newtonian order) are determined by it. Also, in the specific case of the
Schwarzschild spacetime, Dadhich \cite{Da3} has argued how the gravitational
field energy is related to the curvature of space. The above identity implies 
that the horizon is a surface where the magnitude of the gravitational field 
energy equals the gravitational charge. 

The BY quasilocal energy (\ref{BYE}) does not depend on the choice of 
coordinates on the quasilocal two-surface ${\sf B}$. It, however, depends on 
the choice of the foliation and ${\sf B}$ itself. Hence, as we show below, the 
above identity (\ref{Ident}) holds only for specific foliations of such 
spacetimes. 

 

We now prove that Eq. (\ref{Ident}) can be derived from a local relation 
between the covariant derivative of the trace of the extrinsic curvature of 
the timelike three-boundary ${}^3 {\sf B}$ and certain scalars associated with
the Ricci tensor. We first show that this relation is essentially a 
consequence of a Gauss-Codacci embeddability condition on ${}^3 {\sf B}$, 
which is automatically satisfied because of the assumption that the boundary 
${}^3 {\sf B}$ is embeddable in the spacetime. Below, we explicitly state the 
assumptions under which the proof holds. Later we will consider various other
examples of black hole spacetimes, not necessarily static or asymptotically 
flat, for which the identity (\ref{Ident}) is valid.

The first relation we require is the decomposition of the four-dimensional (4D)
Ricci scalar into spatial and timelike components
\be
\label{decomR}
{\cal R} = \g^{\m\n} \g^{\a\b} {\cal R}_{\m\a\n\b} +2n^\m n^\n {\cal R}_{\m\n}
\ \ , \ee
where $\gamma_{\mu\nu}$ is the 3-metric on ${}^3 {\sf B}$ and $n_\mu$ is its
spacelike normal with unit norm. (We shall follow the conventions of Ref.
\cite{MTW}.)
The Gauss-Codacci relation for the projection of the Riemann tensor onto ${}^3 
{\sf B}$ gives for the first term on the right-hand side (rhs) above:
\be
\label{gc1}
\g^{\m\n} \g^{\a\b} {\cal R}_{\m\a\n\b} = R -\T^2 +\T_{\m\n} \T^{\m\n}
\ \ , \ee
where $R$ is the 3D Ricci scalar associated with ${}^3 {\sf B}$ and $\T$ is its
extrinsic curvature. On using the Ricci identity, ${\cal R}_{\m\a\n\b} n^\b = 2
\nabla_{[\m}\nabla_{\a ]} n_\n$, the second term on rhs of Eq. (\ref{decomR}) 
gives
\be
\label{Ric1}
n^\m n^\n {\cal R}_{\m\n} = \T^2 -\T_{\m\n} \T^{\m\n} +\nabla_\m (\T n^\m 
+b^\m ) \ \ ,
\ee
where $b^\m \equiv n^\n \nabla_\n n^\m $. Using Eqs. (\ref{Ric1}) 
and (\ref{gc1}) in the decomposition formula (\ref{decomR}), we obtain 
\be
\label{dTR}
\nabla_\m (\T n^\m +b^\m ) +R = {\cal R} - n^\m n^\n {\cal R}_{\m\n} 
\,. \ee
This is the essential relation we will require below.

We now formulate a condition on the matter distribution that is required for
the identity, Eq. (\ref{Ident}) to hold in asymptotically flat SSS spacetimes. 
Let $\Sigma$ be a smooth hypersurface transverse to the timelike Killing vector
field $t^a$, such that it passes through the bifurcation surface ${\cal H}$ of 
the horizon. Let ${\cal H}_\epsilon$ be a smooth, one parameter family of 
surfaces in $\Sigma$ that approach ${\cal H}$ as $\epsilon\to 0$. We restrict 
our attention to the spacetime region exterior to the (outer) horizon and 
foliate it with a one-parameter family of spacelike hypersurfaces, $\Sigma_t$, 
such that for any value of $t$, the leaf $\Sigma_t$ bears the properties of 
$\Sigma$. Let $u^a$ be a unit timelike normal to $\Sigma$. Then, $t^a = N u^a$,
where $N$ is the lapse function. We shall assume that the following
condition on the Ricci tensor holds in the spacetimes of interest
\be \label{Ricci-flip}
u^\m u^\n {\cal R}_{\m\n} = {\cal R} - n^\m n^\n {\cal R}_{\m\n} \ \ ,
\ee
which is obeyed by the Kerr-Newman family of spacetimes. If we assume that the 
Einstein field equations hold, then this condition translates to
\be \label{Ricci-flipT}
T_{\m\n} \left( u^\m u^\n + n^\m n^\n \right) = 0 \ \ ,
\ee
where $n^\m$ is the unit normal to ${\sf B}$ such that they both lie in $\S$. 
This condition constrains the type of matter allowed to exist in the spacetime 
outside the horizon. The need for such a condition can be understood by the 
fact that the identity (\ref{Ident}) is not expected to hold for additions of 
arbitrary matter field distributions outside the horizon since it can alter 
$E_\infty$.
With this assumption, Eq. (\ref{dTR}) becomes
\be
\label{dTR-flip}
\nabla_\m (\T n^\m +b^\m ) = u^\m u^\n {\cal R}_{\m\n} -R
\,. \ee
Thus, the vector field $[\T n^\m +b^\m ]$ fails to be divergenceless in the 
presence of a non-vanishing rhs, which acts as its source.



We consider spacetimes that are asymptotically flat. This imposes a condition 
on the fall-off behavior of the metric components at spatial 
infinity $i^0$. Note that the metric of any SSS spacetime can be written as
\be \label{SS}
ds^2 = -N^2 dt^2 + \lambda^{-2} dr^2 + r^2 ( d \theta^2 + \sin^2 \theta 
d{\varphi}^2) 
\ \ , \ee
where $N$ and $\lambda$ are dependent on $r$ only. This line element is 
particularly suited to the choice of foliation above, in the sense that $\S_t$ 
is just a $t=$constant hypersurface. Also note that if the two-boundary ${\sf B
}$ is taken to be a sphere of constant curvature-radius $r$, on any $\S_t$, 
then evolving the points on it along the orbits of the Killing field $t^a$ will
generate a three-boundary ${}^3{\sf B}$ that is orthogonal to $\S_t$ wherever
they meet. By asymptotic flatness, we shall mean that $\lambda$ has
the following fall-off behavior: 
\be\label{fall}
\lambda\to 1+\lambda_1r^{-1}+O(r^{-2}) \ \ ,
\ee
as $r\to\infty$. Here, $\lambda_1$ is an $r$-independent constants.

Let $\S_{{\cal H}_\epsilon}$ denote the region on a $\S$-type 
hypersurface that is bounded from the `interior' by ${\cal H}_\epsilon$ and 
from the `exterior' by a two-sphere at infinity. We multiply both sides of Eq. 
(\ref{dTR-flip}) by the lapse $N$ and integrate over the volume of $\S_{{\cal 
H}_\epsilon}$. After a simple rearrangement of terms, we get:
\be\label{T1}
\int_{\S_{{\cal H}_\epsilon}}d^3x \sqrt{-g}\> \nabla_\mu (\Theta n^\mu + b^\mu)
+ \int_{\S_{{\cal H}_\epsilon}}d^3x \sqrt{-g}\>  R = \int_{\S_{{\cal H
}_\epsilon}} d^3 x\> \sqrt{h}\> {\cal R}_{\m\n} u^\m t^\n \ \ ,
\ee
where $h$ is the determinant of the three-metric on $\S_{{\cal H}_\epsilon}$. 
The term on the rhs is the simplest to interpret. When divided by $4\pi$, it 
is just the Komar mass and, therefore, contributes $(E_\infty -M_{{\cal H
}_\epsilon})$, where $M_{{\cal H}_\epsilon}$ is the gravitational charge at 
${\cal H}_\epsilon$. Here, we have implicitly used the fact that $E_\infty
= (M_c )_\infty$. This is justified since at spatial infinity both these 
quantities can be identified with the on-shell Hamiltonian (with the lapse 
tending to unity).

The first term on the left-hand side (lhs) of Eq. (\ref{T1}) yields 
\begin{eqnarray}\label{lhsT}
\int_{\S_{{\cal H}_\epsilon}}d^3x \sqrt{-g}\> \nabla_\mu (\Theta n^\mu + b^\mu)
&=&  \int_{\partial\S_{{\cal H}_\epsilon}}d^2x\>\sqrt{\sigma}\> N\Theta\no \\
&=& \int_{\partial\S_{{\cal H}_\epsilon}} d^2x\>\sqrt{\sigma}\> N K -
\int_{\partial\S_{{\cal H}_\epsilon}}d^2x\>\sqrt{\sigma}\>N n_\mu a^\mu  \ \ ,
\end{eqnarray}
where we used the identity:
\be
\T = \T^t_t + (\T^\theta_{\theta} + \T^\varphi_\varphi ) = -n_\mu a^\mu +K
\ \ ,\ee
which holds here since the spacelike normal $n^\mu$ is orthogonal to $u^\mu$ at
the intersection of $\S$ with ${}^3{\sf B}$ \cite{BY}. Above, $a^\mu \equiv 
u^\n \nabla_\n u^\m$ is the acceleration of the timelike hypersurface normal 
$u^\m$. The result (\ref{lhsT}) is a difference of two terms. The first term is
$8\pi$ times the unreferenced quasilocal mass \footnote{See Eq. (\ref{QM}) 
below for a discussion of this concept.} \cite{BCM,BY} evaluated at infinity 
minus the unreferenced quasilocal mass evaluated near ${\cal H}$. 
The second term, when divided by $4\pi$, is the difference between the Komar 
mass at infinity and the Komar mass at ${\cal H}_\epsilon$. 

The second term on the lhs of (\ref{T1}) can be interpreted as follows. Since 
${}^3 {\sf B}$ is the time evolution of a two-sphere 
embedded in $\S$, its line element is given by Eq. 
(\ref{SS}) with $r=r_0$, where $r_0$ is a constant. Thus, we have $R =2
/r_0^2$ on ${}^3 {\sf B}$. Hence,
\be\label{rk1}
\int_{\S_{{\cal H}_\epsilon}}d^3x \sqrt{-g}\>  R = 8\pi \int_{r_{{\cal H
}_\epsilon}}^\infty dr_0 \>N(r_0) /\lambda(r_0) \,.
\ee
Assuming the integrability of the integrand in the rhs above ensures that 
$N(r) dr/\lambda(r)$ is an exact differential on $\S$, say, 
equal to $df(r)$. Then the above result is equal to
\be\label{rk2}
8\pi\int_{r_{{\cal H}_\epsilon}}^\infty dr \>df(r)/dr =8\pi
f(r)\Big|_{r_{{\cal H}_\epsilon}}^\infty \,.
\ee
On $\S$, suppose $N(r) /\lambda(r)$ approaches unity when $r\to \infty$ and 
when $r\to r_{\cal H}$. Then $f(r) \to r$ in these two neighborhoods. (In fact,
for spherically symmetric electrovac spacetimes, the Einstein field equations
ensure that on $\S$, we have $N(r) =\lambda(r)$ on solutions, for all $r$. 
Hence, the above assumptions are met there.) In such a case, $R =dK_0 /dr$, 
where $K_0 =-2/r$ is the trace of the extrinsic curvature of a two-sphere when 
embedded isometrically (with respect to ${\sf B}$ embedded in $\S$ in the black
hole spacetime) in a {\em flat} spatial slice. 
Then, combining the above results, Eq. (\ref{rk1}) can be reexpressed as 
\be\label{RK}
\int_{\S_{{\cal H}_\epsilon}}d^3x \sqrt{-g}\>  R = 8\pi r\Big|_{r_{{\cal H
}_\epsilon}}^\infty= -\int_{\partial\S_{{\cal H}_\epsilon}} d^2x\> \sqrt{
\sigma} K_0 \,.
\ee
If the spacetime is asymptotically flat, then $1/8\pi$ times the rhs above is 
the appropriate term to be added to the unreferenced Brown-York 
quasilocal energy to obtain the (normalized) physical QLE. 

Using the above results in Eq. (\ref{T1}), we obtain
\be\label{dTR-flip-int0}
{1\over 8\pi} \int_{\partial\S_{{\cal H}_\epsilon}} d^2x\> \sqrt{\sigma} \left(
NK-K_0 \right) = E_{\infty} - M_{{\cal H}_\epsilon} \,.
\ee
In the limit of the interior two-surface, ${\cal H}_\epsilon$, approaching 
arbitrarily close to ${\cal H}$, the above equation implies
\be\label{dTR-flip-int}
\lim_{\epsilon \to 0}{1\over 8\pi} \int_{\partial\S_{{\cal H}_\epsilon}} d^2x\>
\sqrt{\sigma} \left(NK-K_0 \right) = E_{\infty} - M_{\cal H} \ \ ,
\ee
where $M_{\cal H}$ is the Komar mass at ${\cal H}$. The lhs above is the 
difference of contributions from a two-sphere at $r\to \infty$ and from a
two-sphere arbitrarily close to ${\cal H}$. As one approaches ${\cal H}$ along 
$\S$, the lapse $N$ vanishes. For SSS spacetimes (\ref{SS}), even $K=-2\lambda
/r$ vanishes on such a surface. Thus, the contribution from this surface is 
purely due to the reference term $K_0$ and is defined by the following limiting
procedure:
\be\label{BYEH}
E_{\cal H} = \lim_{\epsilon \to 0} E_{{\cal H}_\epsilon}\equiv \lim_{\epsilon 
\to 0} \int_{{\cal H}_\epsilon} d^2 x \sqrt{\sigma} \>{(NK-K_0)\over 8\pi} 
=- \lim_{\epsilon\to 0} \int_{{\cal H}_\epsilon} d^2 x \sqrt{\sigma} \>{K_0
\over 8\pi} \ \ ,
\ee
which is just the QLE at that surface. Let us analyze the contribution from 
the boundary term at spatial infinity. Note that by an earlier assumption we 
have $N/\lambda \to 1$ as $r\to\infty$. Thus, $NK = -2 \lambda^2 / r$ as $r\to 
\infty$. However, $\lambda$ obeys the fall-off condition (\ref{fall}).
Consequently, the contribution to the lhs of (\ref{dTR-flip-int}) from 
infinity is
\be
\lim_{r\to\infty} {1\over 8\pi} \int_{\sf B} d^2x\> \sqrt{\sigma} \left(NK-K_0 
\right) = 2\lambda_1 =2 E_{\infty} \,.
\ee
Hence, the lhs of Eq. (\ref{dTR-flip-int}) is equal to $2E_{\infty} - 
E_{\cal H}$. With these simplifications Eq. (\ref{dTR-flip-int}) itself 
reduces to the required identity, Eq. (\ref{Ident}). 


Note that we have nowhere assumed the spacetime to be a solution of 
general relativity (except in interpreting the condition on the Ricci tensor,
Eq. (\ref{Ricci-flip}), in terms of the constraints posed on the matter stress
tensor). However, the association of $K$ with the quasilocal energy has a 
nice justification \cite{BY,HH} provided the quasilocal
two-surface ${\sf B}$ is taken to be embedded in such a solution. Then, $E$ is 
just the on-shell Hamiltonian with the lapse set equal to 1. 

The spatial volume of integration in Eq. (\ref{T1}) could have been limited to
a different or smaller region of $\S$. In such cases, interpreting the surface 
terms on the lhs of Eq. (\ref{T1}) as QLE is not always possible. In that sense
the horizon and the spatial infinity are very special locations for evaluating 
these terms. 

In the above proof it is assumed that the quantities $E_{\cal H}$, $M_{\cal H}
$, and $E_\infty$, which appear in the identity (\ref{Ident}), are evaluated on
a single leaf, $\S_t$, of the foliation. We now show that the proof can be 
strengthened such that the identity is applicable even in situations where 
contributions from the neighborhood of ${\cal H}$, namely, $E_{\cal H}$ and 
$M_{\cal H}$, are evaluated on a leaf different from the one on which 
$E_\infty$ is computed. To see that this generalization holds, let us construct
the boundary ${}^3{\sf B}$ such that $t^a$ always lies on it. Define
\be\label{Q}
Q_t \equiv {1\over 8\pi}\int_{\sf B} d^2 x \>\sqrt{\sigma} K u^\mu t_\mu \,.
\ee
Then the law of conservation for the matter stress tensor implies \cite{BY}
\be\label{Qcons}
Q_t (\S_{t''} \cap {}^3{\sf B} ) - Q_t (\S_{t'} \cap {}^3{\sf B} )
= - \int_{{}^3{\sf B}} d^3 x \sqrt{-\gamma}\> T^{\mu\nu} n_\mu t_\nu \ \ ,
\ee
which is zero when the source term $T^{\mu\nu} n_\mu t_\nu$
vanishes. In such a case, 
\be\label{QM}
 -Q_t = {1\over 8\pi} \int_{\sf B} d^2 x \sqrt{\sigma} N K  \ \ ,
\ee
which is conserved under diffeomorphisms along the orbits of $t^a$ on  ${}^3{
\sf B}$: it is termed as the quasilocal mass of the system. 
Moreover, when ${\sf B}$ is a sphere of constant curvature-radius in SSS 
spacetimes, then the above equation implies
\be
 -Q_t= N E \ \ ,
\ee
for $N$ is independent of the coordinates on ${\sf B}$. Finally, since the norm
of $t^a$ is fixed along its integral curves, the quasilocal energy $E$ is 
constant on ${}^3{\sf B}$. This completes the generalization of our proof. We 
can, therefore, state our result in the form of the following theorem:

For asymptotically flat, spherically symmetric static solutions of GR, if the 
matter stress tensor obeys Eq. (\ref{Ricci-flipT}) and if, on a constant 
Killing-time hypersurface, the ratio $N(r)/\lambda(r)$ is integrable and
approaches unity at ${\cal H}$ and at $i^0$, then the identity, Eq. 
(\ref{Ident}), is obeyed, and is implied by 
the Gauss-Codacci condition Eq. (\ref{dTR-flip}).
 


We now specifically compute the quasilocal quantities that appear in Eq. 
(\ref{Ident}) for the Reissner-Nordstrom (RN) spacetimes and show that the
identity is obeyed. The corresponding metric and the electromagnetic field can 
be given as
(see, eg., \cite{Carter}):
\begin{mathletters}%
\label{RNsol}
\begin{eqnarray}
ds^2 &=& [C_\t Z_r -C_r Z_\t]\left\{ {dr^2\over\D_r} +{\sin^2\t\over\D_\t}
\right\} \nonumber\\
&&+{\D_\t [C_r dt-Z_r d\varphi]^2 - \D_r [C_\t dt-Z_\t d\varphi]^2
\over [C_\t Z_r -C_r Z_\t]} \ \ ,  \label{RNmet} \\
F&=& {2Q\over r^2}dr\wedge dt -2P\sin\theta d\theta\wedge  d{\varphi}
\ \ , \label{RNem} 
\end{eqnarray}
\end{mathletters}%
where $Q$ and $P$ are the electric and magnetic monopole charges, respectively,
of the hole. Also, $t$ and $r$ are the curvature coordinates.

The electromagnetic stress tensor is
\be\label{emstress}
8\pi T_{ab} = (E^2 + B^2 )\{ \o^{(0)}_a\o^{(0)}_b + 
\o^{(3)}_a\o^{(3)}_b + \o^{(2)}_a\o^{(2)}_b - \o^{(1)}_a\o^{(1)}_b \} \ \ ,
\ee
where $E=Q/r^2$ and $B=P/r^2$ and the tetrad of forms are
\begin{mathletters}%
\label{RNtetrad}
\begin{eqnarray}
\o^{(0)} &=&\sqrt{\Delta_r \over C_\t Z_r -C_r Z_\t }[C_\t dt-Z_\t d\varphi]
\>,\qquad \o^{(1)} =\sqrt{C_\t Z_r -C_r Z_\t \over \Delta_r }dr \ \ ,\\
\o^{(2)} &=&\sqrt{C_\t Z_r -C_r Z_\t \over \Delta_\t }\sin\t d\t\>, \qquad
\o^{(3)} =\sqrt{\Delta_\t \over C_\t Z_r -C_r Z_\t}[C_r dt-Z_r d\varphi]
\,.
\end{eqnarray}
\end{mathletters}%
For RN spacetimes $C_\t =1$, $C_r = 0$, $Z_r = r^2$, $Z_\t = 0$,
$\Delta_r = r^2 -2Mr +Q^2 +P^2$, and $\Delta_\t = \sin^2 \t$,
where $M$ is the mass of the spacetime.

The Ricci tensor is given by
\be\label{RicciRN}
{\cal R}_{ab} = 8\pi T_{ab} \ \ ,
\ee
which is just Einstein's equation for ${\cal R}=0$.
It is clear from Eq. (\ref{emstress}) that 
\be \label{Ricci-flipRN}
\hat{t}^\m \hat{t}^\n {\cal R}_{\m\n} = 
- \hat{r}^\m \hat{r}^\n {\cal R}_{\m\n} \ \ ,
\ee
which is the condition (\ref{Ricci-flip}). Moreover, the metric (\ref{RNmet}), 
when applied to RN spacetimes, has the same form as Eq. (\ref{SS}).
Hence the proof presented for SSS spacetimes remains valid in this case. 
 
The characterizing relation (\ref{Ident}) should hold good in all coordinate
systems for which the spacelike hypersurfaces corresponding to constant 
coordinate-time foliate ${\cal H}$ and spatial infinity in a 
manner identical to the curvature coordinates. Let us in particular verify it 
for an electrically charged hole in isotropic coordinates. The metric in these
coordinates is
 \be \label{RNisomet}
 ds^2 = - \left[\frac {1 - {\alpha^2}/{4r^2}}{1 + {M\over r} + {{\alpha^2}\over
{4r^2}}}\right]^2 dt^2 + (1 + M/r + {\alpha^2}/{4r^2})^2 [dr^2 + r^2 
(d\theta^2 +  \sin^2\theta {d\varphi}^2)] \ \ ,
 \ee
where $\alpha^2 = M^2 - Q^2-P^2$.  For the 
metric (\ref{RNisomet}), the energy and charge expressions are:
 \be \label{EKerr}
 E = M + {\alpha^2}/{2r}
 \ee
  and 
 \be \label{MKerr}
 M_c = \frac {M + {\alpha^2}/r + M{\alpha^2}/{4r^2}}{1 + M/r + 
 {\alpha^2}/{4r^2}}.
 \ee
Hence, the relation (\ref{Ident}) once again gives the (outer) horizon to be at
$r = {\alpha}/2$, showing  that it is valid for isotropic coordinates as well.
 
It is interesting to note that the two cases, $Q = 0 =P$ and $Q^2+P^2 =M^2$, 
characterize conservation  of  charge and  energy, respectively. 
The former is a special case of RN spacetimes; it corresponds to the 
Schwarzschild black hole ($Q=0=P$) spacetime. In this case, the gravitational 
charge obeys $M_c = M$. It is a conserved quantity: the value of $M_c$ is 
independent of the leaf $\S$, or even the location of a two-surface ${\sf B}$ 
(lying outside the horizon) on which it is evaluated.
The latter case describes an extremal RN black hole spacetime. It follows from 
the above expressions for QLE and gravitational charge that the identity 
(\ref{Ident}) holds even for such spacetimes. For an extremal hole it follows 
from Eq. (\ref{EKerr}) that $E = M$ everywhere, implying that it is a conserved
quantity in the above sense. It also means that there is no ``force'' to drive
the collapse and, hence, an extremal hole can never form from the collapse 
of dispersed matter distributions \cite{DN}. 

The difference between gravitational charge and QLE can be appreciated by
considering the particular case of the RN spacetimes. Note that the 
gravitational charge is determined by the ``gravitational charge density''
defined as 
\be
\rho_c = T^0_0 - T^i_i = (1/4 \pi) {\cal R}_{\m\n} u^\m u^\n \ \ , 
\ee
where $u_\m u^\m = -1$ and $i$ is a space index. This should be distinguished
from the matter-energy density, which is $T^0_0 = \rho$. In the case of 
the RN black hole $\rho = (Q^2 +P^2)/2r^4$, while $\rho_c = (Q^2+P^2)/r^4$. 
In the Newtonian approximation, the QLE, $E(r)$, is the sum of matter energy 
density plus the (gravitational and electromagnetic) potential energy required 
to build a ball of fluid by bringing its constituents together from a boundary 
of radius $r$. Furthermore, the contribution to $E(r_0)$ from the region $r>
r_0$ is equal to $(Q^2+P^2)/2r_0$, which is due to the electromagnetic field, 
plus $-M^2/2r_0$, which is due to gravity. Hence the energy enclosed 
by the region is $M-((Q^2+P^2)/2r_0 - M^2/2r_0)$. This is what $E$ is, as 
given by (\ref{EKerr}) (with $r=r_0$ there), in the first approximation.

It would be interesting to explore if the identity (\ref{Ident}) is valid for 
the Kerr-Newman spacetimes, atleast for certain choices of spacetime foliations
and quasilocal two-boundaries. Unfortunately, in this case the exact 
expressions for QLE are not available, except when the two-boundary is taken to
be at spatial infinity. However, Martinez \cite{EAM} has evaluated QLE for
constant stationary time Kerr slices bounded by different types of 
two-boundaries in the {\em slow rotation} approximation. The status of our
identity in this approximate case is being studied \cite{BD}.


In reality a black hole always sits in a cosmological background. It is 
therefore desirable to consider a black hole spacetime that is non-static 
(expanding) and asymptotically FRW. Just as we proved the identity (\ref{Ident}
) generically for asymptotically flat SSS spacetimes earlier, one can similarly
prove it to hold for certain asymptotically non-flat spacetimes as well 
\cite{BD}. Here, we will briefly demonstrate its validity for some of these 
spacetimes. Consider the solution describing an asymptotically FRW, 
electrically and magnetically charged black hole 
\cite{McV,SV}:
\be
\label{isomet}
 ds^2 = - \frac{F^2}{G^2} dt^2 + {S(t)^2}{G^2}H^{-2} 
[dr^2 + r^2 (d\theta^2 + \sin^2\theta d\varphi^2)]
\ee
where
\begin{mathletters}
\begin{eqnarray}
F &=& 1 - H\alpha^2 /(4r^2 S^2) \label{F} \\
H &=& 1 + kr^2 /4, ~\alpha^2 = M^2 - Q^2 -P^2 \label{H} \\ 
G &=& 1 + H^{1/2} M /(rS) + H\alpha^2 /(4r^2 S^2) \label{G} \ \ ,
\end{eqnarray}
\end{mathletters}%
where $k = {\pm 1,0}$, is the space curvature parameter and $S(t)$ is 
the scale factor. Above, we have chosen to write the metric in isotropic
coordinates, which in this case also happens to be comoving.

The QLE in these asymptotically FRW spacetimes is obtained by using Eq. 
(\ref{BYE}), with the corresponding FRW spacetime (where $M=0=Q=P$) chosen as 
the reference. It is evaluated for the foliation comprising of $t=$constant 
hypersurfaces. This gives for the QLE:
\be
E(r) = - Sr^2 G^\prime /H \ \ ,
\ee
in the isotropic gauge. Using the expression for $G$ given in Eq. (\ref{F}), 
we get
\be \label{EbFRW}  
E = MH^{-3/2} + \alpha^2 /(2rSH) \ \ ,
\ee
where $H$ and $\alpha$ are defined above. Here $E({\infty}) = M{H}^{-3/2}$ as 
$rS\rightarrow\infty$. When we switch off the expansion, i.e., set $S = const.
$, we get the energy for an electrically and magnetically charged black hole 
in the Einstein universe. For $S = 1, k = 0$, we 
recover the energy (\ref{EKerr}) in the isotropic coordinates. 
Thus, the above expression has the expected static limit. 

Obtaining the gravitational charge for such non-static spacetimes is more
subtle. Here we adopt a suitable generalization of Eq. (\ref{Mc}). In that 
equation, we identify the lapse function as $N= \sqrt{-g_{tt}}$, where $t$ is 
now the comoving time coordinate in the metric (\ref{isomet}). Such a choice is
motivated by the fact that as $k\to 0$ and $S\to 1$, the above cosmological
metric (\ref{isomet}) approaches the SSS metric (\ref{RNisomet}). Consequently,
the comoving time gets identified with the Killing time of the resulting static
solution. Then, the gravitational charge in such a spacetime is given by
\be
\label{KRN}
M_c = \alpha^2 /2rSH + (MH^{-3/2} + \alpha^2 /2rSH)F /G.
\ee
Note that $M_c (\infty) = E_\infty = MH^{-3/2}$ and Eq. (\ref{Ident}) again 
defines the horizon at $rS = \alpha H^{1/2}/2$. Thus the black hole 
characterization (\ref{Ident}) holds good for an electrically charged black 
hole sitting in an FRW expanding universe. Note that  for $S = 1, k = 0$, we 
recover the gravitational charge (\ref{EKerr}) (with $P=0$) in the isotropic 
coordinates.



Yet another case of applicability of the identity (\ref{Ident}), is an SSS
black hole spacetime with a global monopole charge in it
\cite{BV,HL,Da4}. The metric for the spacetime dual to the Schwarzschild black 
hole incorporating global monopole charge is given by 
 \be \label{globmono}
 ds^2 = -\left(1 - 8 \pi \eta^2 - \frac{2M}{r}\right) dt^2 + \left(1 - 8 \pi 
\eta^2 - \frac{2M}{r}\right)^{-1} dr^2 + r^2 (d \theta^2 + \sin^2\theta d 
\varphi^2) \ \ ,
 \ee
where $\eta$ represents  the  global monopole charge. Setting $\eta = 0$ yields
the Schwarzschild  solution. The event horizon is at $r = 2M/(1 - 8 \pi 
\eta^2)$. The physical effects of global monopole charge  have been studied in 
Refs. \cite{HL,DNY}. The spacetime is not vacuum and is asymptotically non-flat
\cite{Da3}. But it is a solution of the electrogravity-dual vacuum equation 
\cite{Da4}. The 
QLE, on a foliation comprising of $t=$constant hypersurfaces, in the above
spacetime is
  \be
 E = r \left[(1 - 8 \pi \eta^2)^{1/2} - \left(1 - 8\pi \eta^2 - \frac{2M}{r}
\right)^{1/2} \right] \ \ ,
 \ee
where the Hawking-Horowitz prescription \cite{HH} mentioned before was used to
compute the reference contribution of $r (1 - 8 \pi \eta^2)^{1/2}$ above. The 
reference spacetime in this case turns out to be non-flat; rather it is the 
dual-flat solution in the  same sense as the metric (\ref{globmono}) is 
dual-vacuum. Here, $E_\infty = M(1 - 8 \pi \eta^2)^{- 1/2}$, $E_{\cal H} = 2 
E_\infty$. The gravitational charge is $M_c(r\to\infty) = E_\infty$. Again it 
is straightforward to show that Eq. (\ref{Ident}) will define the location of 
the horizon. 

In our discussions above of black hole solutions that are asymptotically FRW
or that have a global monopole charge, we have demonstrated that our identity 
(\ref{Ident}) is applicable to non-static as well as asymptotically non-flat
cases. The details of the derivation of this identity for such spacetimes from 
a requirement of the type (\ref{dTR}), will be given elsewhere \cite{BD}.

   
Note that the identity (\ref{Ident}) has the following implication on the 
non-attainment of extremality. The  particular expression for the gravitational
charge of a black hole, $M_{\cal H} = (\kappa/4 \pi)A$, relates it 
to the surface gravity $\kappa$ (and, therefore, the temperature) of the hole. 
Here $A$ is area of the 
horizon. The third law of  black  hole dynamics  states  that it is impossible 
to  reduce  gravitational charge of a hole to zero by a finite  sequence  of  
physical processes \cite{DN}. In view of the relation (\ref{Ident}), we could 
as well say that the magnitude of the field energy, $| E_\infty -E_{\cal H}|$
cannot be reduced to zero in  a  finite sequence  of  physical interactions. 
Since the surface gravity of the RN hole is zero in the extremal limit, i.e.,
$M^2 = Q^2 +P^2$, the field energy is also zero in this case, which implies 
that an extremal hole can never be formed from the collapse of dispersed 
matter distributions. Similarly, a non-extremal hole can never turn extremal, 
say, due to infalling charged matter. Recent quantum field
theoretic and topological considerations seem to suggest that the
converse  may also be true, i.e., extremal RN holes may also be prevented from
turning into non-extremal ones (see Refs. \cite{HH,CT}). In that case the 
extremal and non-extremal holes would be analogous to particles of zero and 
non-zero  mass, respectively, where the gravitational charge acts as a nice 
analogue of mass \cite{Da2}.

We observe that the {\em horizon}-based quantity $(E_{\cal H} - M_{\cal H})$ is
the analogue of the internal energy of the whole spacetime (up to the addition 
of an exact form) in the thermodynamical laws of static black holes. This is 
because $E_\infty$ itself has the interpretation of being the thermodynamical
internal energy of the whole spacetime \cite{BY}.

It was known from the formulation of the laws of black hole mechanics that {\em
variations} of certain quantities at the horizon and at infinity are related. 
Here, we have shown that a non-variational identity relating some of these 
quantities also exists in general relativity. 
In the past, Iyer and Wald have also explored similar possibilities
\cite{IW}. These authors generalized the definition of the BY quasilocal mass 
to a more general class of diffeomorphism invariant Lagrangian theories of
gravity. One of the results of their work is that as one approaches close to 
${\cal H}$ along a smooth hypersurface $\S$, which is 
transverse to the Killing time $t^a$, the Noether charge associated with $t^a$ 
approaches twice the boundary terms in the gravitational action (which in turn 
depend on the choice of boundary conditions imposed on the dynamical fields). 
Our identity is similar in spirit to (but different in content from) this 
relation. We are currently studying the possibility of an identity similar 
to Eq. (\ref{Ident}) existing in similar Lagrangian theories of gravity.

Finally, it is always welcome to gain some insight into the difficult  and  
ambiguous concept of energy in GR.  Most  of  the definitions  refer to a 
quasilocal energy, which generally  includes contribution of the matter energy 
density, while  gravitational charge is essentially defined through the Komar 
integral or its generalizations. The latter is related to the formulation of 
the Gauss law  for  stationary spacetimes \cite{DC}. Here we have extended the 
application of these constructs to non-static and asymptotically non-flat 
spacetimes as well. Although we demonstrated that our identity (\ref{Ident}) is
applicable to these cases, after suitably adapting the Komar integral 
(\ref{Komar}) for such
examples, we did not derive it from an ``embeddability'' condition of 
the type (\ref{dTR}). Moreover, even where we proved the identity, our 
consideration was limited essentially to eternal black hole spacetimes, where 
the apparent and event horizons overlapped. To be astrophysically relevant, 
however, one must deal with the case of isolated horizons \cite{ABF}. 
Generalization of the proof for the applicability of an identity of the type 
(\ref{Ident}), to other black hole spacetimes, and to the case of 
isolated horizons, is presently under consideration \cite{BD}.


We thank Abhay Ashtekar for helpful discussions. We also thank the referee for
valuable remarks. One of us (ND) would like to thank the Commission for 
Universities and Research of the Government of Catalunya for the award of a 
visiting fellowship and the University of Barcelona (Department of Fundamental 
Physics) for their hospitality.

  \end{document}